\documentstyle[epsfig,12pt]{article}

\newcommand{\bce}{\begin{center}} 
\newcommand{\ece}{\end{center}}
\newcommand{\beq}{\begin{equation}}
\newcommand{\eeq}{\end{equation}}
\newcommand{\bea}{\vspace{0.25cm}\begin{eqnarray}}
\newcommand{\eea}{\end{eqnarray}}

\newcommand{\br}{{\bf r}}

\newcommand{\ba}{\begin{array}}
\newcommand{\ea}{\end{array}}

\newcommand{\ket}[1]{| {#1} \rangle}
\newcommand{\bra}[1]{\langle {#1} |}

\newcommand{\doublespace}{
    \renewcommand{\baselinestretch}{1.6}\large\normalsize}

\def\lsim{\mathrel{\rlap{\lower4pt\hbox{\hskip1pt$\sim$}}
    \raise1pt\hbox{$<$}}}         
\def\gsim{\mathrel{\rlap{\lower4pt\hbox{\hskip1pt$\sim$}}
    \raise1pt\hbox{$>$}}}         

\def\beq{\begin{equation}}
\def\endeq{\end{equation}}
\def\arr{\begin{eqnarray}}
\def\endarr{\end{eqnarray}}
\makeindex

  \advance\oddsidemargin  by -1in
  \advance\evensidemargin by -1in
  \advance\topmargin      by -0.5in
\begin{document}

\vspace{2.0cm}

\begin{flushright}
\end{flushright}

\vspace{1.0cm}

\begin{center}
{\Large \bf 
Higher twist effects  in charmed-strange  $\nu$DIS diffraction} 

\vspace{1.0cm}

{\large\bf R.~Fiore$^{1 \dagger}$ and V.R.~Zoller$^{2 \ddagger}$}

\vspace{1.0cm}

$^1${\it Dipartimento di Fisica,
Universit\`a     della Calabria\\
and\\
 Istituto Nazionale
di Fisica Nucleare, Gruppo collegato di Cosenza,\\
I-87036 Rende, Cosenza, Italy}\\
$^2${\it
ITEP, Moscow 117218, Russia\\}
\vspace{1.0cm}
{ \bf Abstract }\\
\end{center}
  The non-conservation of charmed-strange   current in 
 the  neutrino deep inelastic scattering ($\nu$DIS)
strongly affects the longitudinal 
structure function, $F_L$, at small values of Bjorken $x$.
 The corresponding  correction to $F_L$   is a  higher twist effect
 enhanced at small-$x$ by the  rapidly growing  gluon density factor.
As a result, the component of $F_L$ induced by the charmed-strange
current prevails over the light-quark component and 
 dominates $F_L=F_L^{cs}+F_L^{ud}$ at $x\lsim 0.01$ and $Q^2\sim m_c^2$. 
 The color dipole analysis  clarifies the physics behind the phenomenon
and provides a quantitative estimate of the effect.

\doublespace

\vskip 0.5cm \vfill $\begin{array}{ll}
^{\dagger}\mbox{{\it email address:}} & \mbox{fiore@cs.infn.it} \\
^{\ddagger}\mbox{{\it email address:}} & \mbox{zoller@itep.ru} \\
\end{array}$

\pagebreak



\section{Introduction}
Weak currents are not conserved.  For the  light flavor currents  
the hypothesis of the partial conservation
 of the axial-vector current (PCAC) \cite{PCAC} provides quantitative measure
of the   
charged current  non-conservation (CCNC) effect.
\cite{Adler}.  The  non-conservation of the charm and strangeness changing 
($cs$) current  is not 
constrained by PCAC. Here we focus
on manifestations of the $cs$ 
current  non-conservation
in small-x neutrino DIS. At small $x$ the color dipole (CD) approach to QCD
\cite{NZ91,M}
proved to be very effective. Within this approach  it is natural to 
quantify the effect of CCNC
 in terms of 
the light cone wave functions (LCWF)\footnote{Preliminary results have 
been reported at the Diffraction'08 Workshop \cite{FZCS3}}
\beq
\Psi\sim g\epsilon_{\nu}j^{\nu}
/\Delta E,
\label{eq:Psi}
\eeq 
where $j_{\nu}=\bar c(k) 
\gamma_{\nu}(1-\gamma_5)s(p)$, $\Delta E=E_q-E_p-E_k$ and   
$\epsilon_{\nu}$ is the four-vector 
of the so-called longitudinal polarization 
of the  $W$-boson with the  four-momentum $q$. Notice that
$\epsilon_{\nu}\to q_{\nu}/Q$ for $Q^2=-q^2\to 0$.

The observable which is   highly  
sensitive to the 
CCNC effects is  the 
longitudinal structure function 
$F_L(x,Q^2)$ related, within the CD approach ,  to
the quantum mechanical expectation value of the color dipole cross section,
\beq
F_L\sim Q^2\bra{\Psi}\sigma\ket{\Psi}.
\label{eq:FLCD}
\eeq  
Our finding is that
the  higher twist  correction to $F_L$ arising  from  the 
$cs$ current non-conservation  appears to be  
enhanced at small $x$ by the BFKL \cite{BFKL} gluon density 
factor,
\beq
F_L^{cs} \sim {m_c^2\over Q^2}\left(1\over x\right)^{\Delta}.
\label{eq:HT}
\eeq 
The color dipole analysis reveals  mechanism of enhancement:
the  ordering of 
dipole sizes 
$$(m_c^2+Q^2)^{-1}<r^2<m_s^{-2}$$
typical of the Double 
Leading Log Approximation (DLLA) and  
the  multiplication of  $\log$'s 
like 
$$\alpha_S\log[(m_c^2+Q^2)/\mu^2_G)\log(1/x) $$
to
higher orders 
of perturbative QCD. As a result, the component $F_L^{cs}$ 
induced by 
the charmed-strange 
current 
\beq
F_L=F^{ud}_L+F^{cs}_L
\label{eq:FLUDCS}
\eeq
grows rapidly to small-$x$ and 
 dominates $F_L$ at   $Q^2\lsim m_c^2$ \cite{FZCS1,FZCS2}.

\section{CCNC in terms of LCWF }
In the CD approach to small-$x$ 
$\nu$DIS \cite{Kolya92} the responsibility for the quark current
 non-conservation
 takes the light-cone wave function  of the quark-antiquark
Fock state of the longitudinal ($L$)  electro-weak 
boson\footnote{for an alternative description of the   $\nu$DIS 
structure functions see e.g. \cite{Kretzer}.}. 
For  Cabibbo-favored transitions  
the Fock state  expansion   reads
\beq
\ket{W^+_L}=\Psi^{cs}\ket{c\bar s}+\Psi^{ud}\ket{u\bar d}+...,
\label{eq:FOCK}
\eeq  
where only   $u\bar d$- and $c\bar s$-states (both vector and axial-vector) 
are retained.
 
In  the current conserving eDIS the Fock state expansion  of 
the longitudinal photon
   contains only $S$-wave $q\bar q$ states and $\Psi$  vanishes as  
$Q^2\to 0$,
\beq
\Psi(z,{\bf r})\sim 
2\delta_{\lambda,-\bar\lambda}Q z(1-z)\log(1/\varepsilon r).
\label{eq:V1}
\eeq
Here $\bf r$ is the $q\bar q$-dipole size and $z$ stands for the 
Sudakov variable of the quark.

In $\nu$DIS the CCNC  adds to Eq.(\ref{eq:V1})
 the $S$-wave  mass term \cite{FZ1,FZ2}
\beq
\sim \delta_{\lambda,-\bar\lambda}Q^{-1}
\left[(m\pm \mu)[(1-z)m\pm z \mu]\right]\log(1/\varepsilon r)
\label{eq:V2}
\eeq
and generates the $P$-wave component of $\Psi(z,{\bf r})$, 
\beq
\sim i\zeta\delta_{\lambda,\bar\lambda}e^{-i2\lambda\phi}Q^{-1}(m\pm \mu)r^{-1},
\label{eq:P1}
\eeq
 where upper sign is  for the  axial-vector current, lower - for the vector one and 
$\zeta=2\lambda$ -
for the vector current and $\zeta=1$ - for the axial-vector one.
 Clearly seen are the built-in
divergences of
the vector and axial-vector currents
$\partial_\nu V^{\nu}\sim m-\mu$ and  $\partial_\nu A^{\nu}\sim m+\mu$. 
This LCWF describes the quark-antiquark state with  quark  of mass  
$m$ and  helicity   $\lambda=\pm 1/2$
carrying fraction $z$ of the $W^+$ light-cone momentum and 
antiquark having  mass $\mu$, helicity   $\bar\lambda=\pm 1/2$
\begin{figure}[h]
\psfig{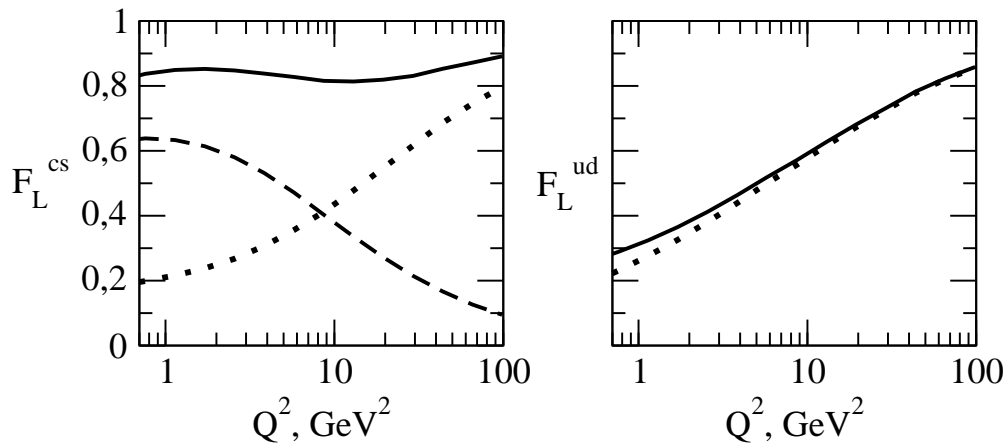}
\vspace{-0.5cm}
\caption{Two components of 
$F_L=F^{cs}_L + F^{ud}_L$ at $x_{Bj}=10^{-4}$ are shown 
by solid lines. The $S$-wave and $P$-wave contributions to $F^{cs}_L$  and 
$F^{ud}_L$ are represented by dotted and dashed lines, correspondingly.} 
\label{fig:fig1}
\end{figure} 
 and 
momentum fraction $1-z$.
The distribution of  dipole sizes is controlled by the 
attenuation parameter
$$\varepsilon^2=Q^2z(1-z)+(1-z)m^2+z\mu^2$$ 
that introduces, in fact, the 
infrared cut-off,
$r^2\sim\varepsilon^{-2}$.  

\section{High $Q^2$: $z$-symmetric $c\bar s$-states}
In the color dipole representation \cite{NZ91,M}
the  longitudinal structure 
function $F_L(x,Q^2)$ 
in  the vacuum exchange dominated  region
of $x\lsim 0.01$ 
can be represented in a factorized form
\beq
F_{L}(x,Q^{2})
={Q^2\over 4\pi^2\alpha_{W}}\int dz d^{2}{\bf{r}}
|\Psi(z,{\bf{r}})|^{2} 
\sigma(x,r)\,,
\label{eq:FACTOR}
\eeq
where $g$ is the weak charge,  $\alpha_W=g^2/4\pi$ and 
${G_F/\sqrt{2}}={g^2/m^2_{W}}$. The light 
cone density of 
color  dipole states  $|\Psi|^2$ 
is the incoherent sum of the vector $(V)$  and the  
axial-vector $(A)$ terms, 
\beq
|\Psi|^2= |V|^2+ |A|^2.
\label{eq:PSI2}
\eeq

The Eq. (\ref{eq:V1}) makes it obvious
that  for large enough virtualities of the probe, $Q^2\gg m_c^2$,
   the $S$-wave  components of both $F^{ud}_L$ and $F^{cs}_L$ in expansion 
(\ref{eq:FLUDCS})
 are dominated by the ``non-partonic'' 
configurations with  $z\sim 1/2$  with  characteristic dipole sizes
\cite{BGNPZ94} 
$$r^2\sim Q^{-2}.$$ 
In the CD  approach the BFKL-$\log(1/x)$ evolution \cite{BFKL} 
of  $\sigma(x,r)$ in Eq.(\ref{eq:FACTOR})
is described by the CD BFKL equation of Ref.\cite{NZZBFKL}. 
For qualitative estimates it suffices to use  
the DLLA
(also known as  DGLAP approximation \cite{D,GLAP})
Then, for  small dipoles \cite{NZFL}  
\beq
\sigma(x,r)
\approx {\pi^2 r^2\over N_c}\alpha_S(r^{-2})
G(x,r^{-2}),
\label{eq:SMALL}
\eeq
and from Eq.(\ref{eq:FACTOR}) it follows that
\beq
F^{ud}_L\approx F^{cs}_L\approx {2\over 3\pi}\alpha_S(Q^2)G(x,Q^2),
\label{eq:FL2FL}
\eeq
where $G(x,k^2)=xg(x,k^2)$ is the 
gluon structure function and 
$\alpha_S(k^2)=4\pi/\beta_0\log(k^2/\Lambda^2)$ with
$\beta_0=11-2N_f/3$.

The $rhs$ of (\ref{eq:FL2FL})  is quite similar 
 to  $F^{(e)}_L$ of eDIS \cite{D,Cooper} 
(see \cite{NZFL} for discussion of corrections to 
DLLA-relationships 
between the gluon density $G$ and $F^{(e)}_L$). 
Two $S$-wave terms in the   
 expansion (\ref{eq:FLUDCS}) 
that mimics the expansion (\ref{eq:FOCK}) evaluated within the CD BFKL model 
of Ref.\cite{NSZZ}  are 
shown by dotted curves in  Fig.~\ref{fig:fig1}. The  full scale BFKL 
evolution of 
the $\nu N$ structure function $F_L(x,Q^2)$ with boundary 
condition at $x_0=0.03$ 
is shown in Fig. 2 of Ref.\cite{FZAdler}.

\section{Moderate $Q^2$: asymmetric $c\bar s$-states and $P$-wave dominance}
The $S$-wave term dominates $F_L$  at high 
$Q^2\gg m_c^2$.
At moderate $Q^2\lsim m_c^2$ the $P$-wave component takes over 
(see Fig.{\ref{fig:fig1}}). To  evaluate it 
we turn to  Eq. (\ref{eq:FACTOR}).
For   $m_c^2\gg m_s^2$ in Eq.(\ref{eq:PSI2}), 
$$
|V_L|^2\sim|A_L|^2\propto
\left({m_c^2\over Q^2}\right)
\varepsilon^2 K^2_1(\varepsilon r)$$,
where $K_1(x)$ is the modified Bessel function 
and one can integrate in (\ref{eq:FACTOR})  over $r^2$ to see that
the  $z$-distribution, $dF^{cs}_L/dz$,   develops  
the parton model peaks at $z\to 0$ and $z\to 1$ \cite{FZCS1}. 
To clarify the issue of relevant dipole sizes we
integrate in (\ref{eq:FACTOR})  first over $z$   near the endpoint $z=1$.
For $r^2$ from the region 
$$
(m_c^{2}+Q^2)^{-1}\lsim r^2 \ll m_s^{-2}$$ this 
yields \cite{FZCS2}
\bea
\int dz |\Psi^{cs}(z,\br)|^2\approx {\alpha_WN_c\over \pi^2}{m_c^2\over
 (m_c^2+Q^2)}{1\over Q^2r^4}.
\label{eq:ZINT1}
\eea

This is the $r$-distribution for $c\bar s$-dipoles with $c$-quark
 carrying a fraction 
$z\sim 1$ of 
the  $ W^+$'s  light-cone momentum. Thus, the singularity $\sim r^{-4}$
in Eq.(\ref{eq:ZINT1}) together with the factorization relation 
(\ref{eq:FACTOR}) and $\sigma(r)\sim r^2$ 
give rise to nested logarithmic integrals over dipole sizes.
 Indeed, 
in the Born approximation the gluon density $G$ in Eq.(\ref{eq:SMALL}) is   
\beq
G(x,r^{-2})\approx C_FN_cL(r^{-2}),
\label{eq:2G}
\eeq
where 
\beq
L(k^2)={4\over \beta_0}
\log{\alpha_S(\mu_G^2)\over\alpha_S(k^2)}.
\label{eq:L}
\eeq
Notice, that perturbative gluons do not propagate to large distances
 and $\mu_G$ in Eq.(\ref{eq:L})
stands  for the inverse Debye screening radius, $\mu_G=1/R_c$.
The lattice QCD data suggest $R_c\approx 0.3$ fm \cite{MEGGIO}.
Because $R_c$    is small  compared to the typical range of strong 
interactions, the dipole cross section  evaluated with  the  decoupling of 
soft gluons, $k^2\lsim \mu_G^2$,
 would underestimate  the interaction strength for
 large color dipoles. In Ref.\cite{NPT,NSZZ,LANACH} this missing strength
 was modeled by 
a non-perturbative, soft correction $\sigma_{npt}(r)$ to the 
dipole cross section $\sigma(r)=\sigma_{pt}(r)+\sigma_{npt}(r).$ 
Here we concentrate on  the perturbative component, $\sigma_{pt}(r)$,
represented by Eqs.(\ref{eq:SMALL}) and (\ref{eq:2G}).

Then, for the charmed-strange $P$-wave component of  $F_L$ with fast $c$-quark
($z\to 1$) one gets
\beq
F_L^{cs}\sim 
{N_cC_F\over 4}{m_c^2\over {(m_c^2+Q^2)}}{1\over {2!}}L^2(m_c^2+Q^2).
\label{eq:FL2G}
\eeq
There is also a  contribution to 
$F_L^{cs}$  from the region $0<r^2<(m_c^2+Q^2)^{-1}$
\bea  
F_L^{cs}\sim 
{N_cC_F\over 4}{m_c^2\over {(m_c^2+Q^2)}}\alpha_S(m_c^2+Q^2)L(m_c^2+Q^2)
\label{eq:FLPEN1}
\eea
which is, however  one  $L$ short.
\begin{figure}[t]
\rule{5cm}{0.2mm}\hfill\rule{5cm}{0.2mm}
\vskip 2.5cm
\rule{5cm}{0.2mm}\hfill\rule{5cm}{0.2mm}
\psfig{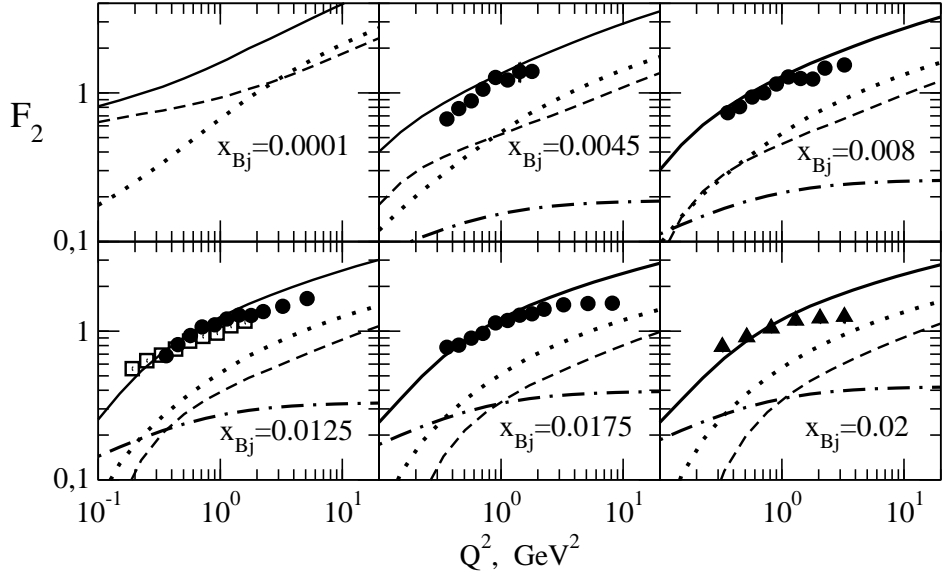}
\caption{The nucleon structure function $F_2$ at smallest available  
$x_{Bj}$ as measured in   $\nu Fe$ CC DIS by the 
CCFR 
\cite{Fleming} 
(circles) and  CDHSW Collaboration 
\cite{CDHSW} 
(squares, $x_{Bj}=0.015$). 
Triangles are  the 
CHORUS Collaboration  measurements 
\cite{CHORUS}
of $F_2$  in 
 $\nu Pb$ CC DIS. 
 Solid curves show the vacuum  exchange 
contribution to $F_2$.
Also shown are the charm-strange (dashed curves) and light 
flavor (dotted curves) components of $F_2$, dashed-dotted curves for the valence 
contribution to $F_2$.
\label{fig:fig2}}
\end{figure}  
Thus,
 the CD   analysis  reveals the
 ordering of dipole sizes
\beq 
(m_c^{2}+Q^2)^{-1}\lsim r^2 \ll m_s^{-2}
\label{eq:ORDER}
\eeq
typical of the DGLAP approximation. 
The rise of $F^{cs}_L(x,Q^2)$ towards small $x$ is generated by interactions
of the higher Fock states,  $c\bar{s}+gluons$.
The DLLA ordering of Sudakov variables and dipole sizes in  the $n$-gluon state
$\ket{c\bar{s}g_1g_2...g_n}$
\beq 
x\ll z_n\ll...\ll z_1\ll z < 1
\label{eq:ZORDER}
\eeq

\beq 
(m_c+Q^2)^{-1}\ll r^2\ll \rho_1^2\ll...\ll \rho^2_n\ll \mu_G^{-2}
\label{eq:RORDER}
\eeq
results in the density $|\Phi_{n+1}|^2$ of multi-gluon
states in the color dipole space \cite{NZ91}
\bea
|\Phi_{n+1}|^2=|\Psi(z,{\bf{r}})|^{2}{C_F\alpha_S(r^{-2})\over \pi^2}
\cdot{1\over z_1}\cdot
{r^2\over\rho_1^4 }\nonumber\\
\times {C_F\alpha_S(\rho_1^{-2})\over \pi^2}\cdot{1\over z_2}\cdot
{\rho_1^2\over\rho_2^4}\dots
{C_F\alpha_S(\rho_{n-1}^{-2})\over \pi^2}\cdot{1\over z_n}\cdot
{\rho_{n-1}^2\over\rho_n^4}.
\label{eq:PhiN}
\eea
By virtue of (\ref{eq:ZORDER},\ref{eq:RORDER}) 
the $c\bar{s}g_1g_2...g_n$-state interacts like color singlet octet-octet 
state with the cross section $(C_A/C_F)\sigma(\rho_n)$. 
Then, making an explicite use of Eqs.(\ref{eq:SMALL},\ref{eq:2G}) and 
(\ref{eq:ZINT1}) 
we arrive at
the $P$-wave component of $F_L$  that rises rapidly to small $x$,
\bea
F_L^{cs}\approx \left(Q^2\over 4\pi^2\alpha_{W}\right)
\pi^2C_F\int dz d^{2}{\bf{r}}|\Psi(z,{\bf{r}})|^{2}\nonumber\\
\times r^2\alpha_S(r^{-2})\sqrt{L(r^{-2})\over \eta}
I_1\left(2\sqrt{\xi(x,r^{-2})}  \right)\nonumber\\
\approx  {N_cC_F\over 4}{m_c^2\over {(m_c^2+Q^2)}}{L(m_c^2+Q^2)\eta^{-1}}
I_2\left(2\sqrt{\xi(x,m_c^2+Q^2)}\right).
\label{eq:DGLAP1}
\eea
In Eq.(\ref{eq:DGLAP1}),  which is the DGLAP-counterpart of Eq.(\ref{eq:HT}), 
$$I_{1,2}(z)\simeq \exp(z)/\sqrt{2\pi z}$$ 
is the Bessel function, 
$$\xi(x,k^2)= \eta L(k^2)$$ is the DGLAP expansion parameter with
$
\eta=C_A\log(x_0/x).$

Additional contribution to $F^{cs}_L$ 
comes from the $P$-wave $c\bar s$-dipoles 
with ``slow'' $c$-quark,  $z\to 0$.
 For low $Q^2\ll m_c^2$ this contribution is rather small,
\beq
F_L^{cs}\approx {N_cC_F\over 4}
{(Q^2+m_s^2)\over m_c^2}\left(\alpha_S^2\over\pi\right)^2\log(m_c^2/\mu^2_G).
\label{eq:FLZ0}
\eeq
If, however,  $Q^2$ is large enough, $Q^2\gg  m_c^2$,
corresponding distribution of dipole sizes  valid for 
$$ 
(m_c^{2}+Q^2)^{-1}\lsim r^2 \ll m_c^{-2}$$  
is
\bea
\int dz |\Psi^{cs}(z,\br)|^2\approx {\alpha_WN_c\over \pi^2}{m_c^2\over
 Q^2}{1\over Q^2r^4}.
\label{eq:ZINT2}
\eea
The DLLA summation over the $s$-channel  multi-gluon  states, 
results in \cite{FZCS3}
\bea
F_L^{cs}\approx  {N_cC_F\over 4}{m_c^2\over {Q^2}}{L(Q^2)\eta^{-1}}
I_2\left(2\sqrt{\xi(x,Q^2)}\right).
\label{eq:DGLAP2}
\eea
Therefore, at high $Q^2\gg m_c^2$ both  kinematical 
domains  $z\to 1$ and $z\to  0$ (Eqs.(\ref{eq:DGLAP1}) and (\ref{eq:DGLAP2}), 
respectively)
contribute (within the DLLA accuracy) equally to $F_L^{cs}$.

\section{Low $Q^2$: light quark  dipoles and Adler's theorem.}
 The P-wave component of $F_L^{ud}$  is small
because of  small factor  $m^2_q/Q^2$, where $m_q$ is the 
constituent $u,d$-quark mass. Here we deal with constituent quarks  
in the spirit of  Weinberg \cite{Weinberg}.  
This suppression factor, $m^2_q/Q^2$,
comes from the light-cone wave 
function $\Psi^{ud}\sim m_q(Qr)^{-1}$ and is of purely perturbative nature.

In \cite{FZAdler}
we checked accuracy  of 
the color dipole description of $F_L(x,Q^2)$  
in the non-perturbative domain of low $Q^2$ making 
use of   Adler's theorem \cite{Adler},
\beq
F^{ud}_L(x,0)= {f^2_\pi\over \pi}\sigma_\pi,
\label{eq:Adler}
\eeq
In  (\ref{eq:Adler}) $f_{\pi}$ is the pion decay constant, $\sigma_\pi$ is 
 the on-shell pion-nucleon total cross section.

 Invoking the CD factorization,
which is valid for  soft as well as for hard
diffractive interactions,
 we evaluated first the vacuum exchange contribution to both
 $\sigma_{\pi}$ and  $F_L(x,0)$.   
The parameter $f_{\pi}$ in 
 Eq.(\ref{eq:Adler}) was evaluated 
within  the CD LCWF technique \cite{Jaus,SNS}.
The approach successfully passed the consistency test:
$\pi F^{ud}_L(x,0)/(f^2_\pi\sigma_\pi)\approx 1$ to within $10\%$. 
The cross section $\sigma_{\pi}$ was found to be in agreement with data.
However, the value of  
$f_{\pi}$ appeared to be underestimated.
It was found that for  $m_q=150$ MeV,   commonly used now 
in CD models successfully tested against DIS data, 
our $F_L$ at $Q^2\to 0$ undershoots the empirical value of 
$f^2_\pi\sigma_\pi/\pi$
by about   $40\%$ \cite{FZAdler}, not quite bad
for the model evaluation of  non-perturbative parameters. One can think 
of  improving  accuracy  at  higher $Q^2\sim m_c^2$ which we are interested
 in.

Notice,  that Adler's theorem
allows only a slow rise of $F^{ud}_L(x,0)$ to small $x$, 
\beq
F^{ud}_L(x,0)  \propto \left(1\over x\right)^{\Delta_{soft}},
\label{eq:008}
\eeq
much slower than the rise  of $F_L^{cs}$ following from our  DLLA estimates.
The value of the  so-called soft pomeron intercept  $\Delta_{soft}\simeq 0.08$ 
comes from the Regge parameterization of the total $\pi N$ cross section 
\cite{DonLan}.

\section{Comparison with experimental data.}
We evaluate nuclear $(\nu A)$ and nucleon $(\nu N)$ structure functions  within the 
color dipole BFKL approach  \cite{NSZZ}
(for alternative approaches to nuclear shadowing in neutrino DIS 
see \cite{MILTHOM,BROD,QIU,MACHADO,KULPET}). 

The structure function $F_2$ 
for the $\nu Fe$ and $\nu Pb$ interactions are  shown
in Fig. {\ref{fig:fig2}. From  
  comparison  with 
experimental data \cite{Fleming}, \cite{CDHSW} and \cite{CHORUS}  
we conclude that the excitation of charm 
contributes significantly to $F_2$ at $x\lsim 0.01$ and dominates $F_2$
at $x\lsim 0.001$ and $Q^2\lsim m_c^2$. 

 For comparison with data taken at moderately small-$x$ the valence component,
 $F_{2\,val}$,  
of the structure function $F_2$ should be taken into account.
We resort to the parameterization of  $F_{2\,val}(x,Q^2)$ suggested in \cite{Reno}. 
This parameterization 
gives $F_{2\,val}(x,Q^2)$ vanishing as $Q^2\to 0$ which  is not 
quite satisfactory from the point of view of PCAC. 
The latter requires
$F_{2\,val}(x,0)=F^{PCAC}_{2\,val}(x,0)$ with
$$F^{PCAC}_{2\,val}(x,0)={f^2_{\pi}\over \pi}\sigma^R_{\pi}(W).$$  Here  
$x=m_a^2/W^2$ and $\sigma^R_{\pi}(W)$ stands for the secondary reggeon 
contribution to the total pion-nucleon cross section that diminishes  at high 
{\it cms} collision energy as $\sigma^R_{\pi}(W)\sim (W^2)^{\alpha_R-1}$, 
where $\alpha_R\simeq 0.5$. However, at smallest values of 
$Q^2\simeq 0.2-0.3$ GeV$^2$ accessible 
experimentally 
$F_{2\,val}(x,Q^2)\gg F^{PCAC}_{2\,val}(x,0)$, remind, the  characteristic mass scale 
in the axial channel is $m_a\sim 1$ GeV. Therefore,
the accuracy
of $F_{2\,val}(x,Q^2)$ of Ref.\cite{Reno} is quite sufficient 
for our purposes. 
In  Fig. {\ref{fig:fig2}  the valence contributions to $F_2$  are 
 shown by  dash-dotted curves. The agreement with 
data is quite reasonable.

One more remark is in order, the perturbative 
light-cone density of $u\bar d$ states, 
$|\Psi^{ud}|^2\sim r^{-2}$, apparently overestimates the role  of 
short distances at low  $Q^2$ (see Ch. 5)  and gives the value of  
$F_L^{ud}(x,0)$ 
which is 
smaller than the value dictated  by Adler's theorem  \cite{FZAdler}.
This also  may lead to underestimation of  $F_2$ in the region of 
moderately  small $x\gsim 0.01$ dominated by the light quark current.

\section{Summary} 
 
Summarizing, it is shown  that at small $x$ and moderate   
virtualities of the probe, $Q^2\sim m_c^2$, the higher twist corrections
brought about by  the non-conservation  of the  charmed-strange current 
dramatically change the longitudinal structure function, $F_L$.
   The effect   survives 
the limit $Q^2\to 0$ and  
seems to be interesting from a point of view of 
 feasible  tests of Adler's theorem \cite{Adler} and the PCAC hypothesis.

\section{Acknowledgments}

V.R.~Z. thanks  the Dipartimento di Fisica dell'Universit\`a
della Calabria and the Istituto Nazionale di Fisica
Nucleare - gruppo collegato di Cosenza for their warm
hospitality while a part of this work was done.
The work was supported in part by the Ministero Italiano
dell'Istruzione, dell'Universit\`a e della Ricerca and  by
 the RFBR grants 07-02-00021 and 09-02-00732.

\end{document}